\documentstyle[multicol,aps,prl,epsf]{revtex}
\begin{document}

\draft

\title{
Spin-triplet superconductivity in 
repulsive Hubbard models with disconnected Fermi surfaces: 
a case study on triangular and honeycomb lattices
}

\author{
Kazuhiko Kuroki and Ryotaro Arita
}
\address{Department of Physics, University of Tokyo, Hongo,
Tokyo 113-0033, Japan}

\date{\today}

\maketitle

\begin{abstract}
We propose that spin-fluctuation-mediated 
spin-triplet superconductivity may be  
realized in repulsive Hubbard models with disconnected 
Fermi surfaces. The idea is confirmed for Hubbard models on
triangular (dilute band filling) and honeycomb (near half-filling) 
lattices using fluctuation exchange 
approximation, where triplet pairing order parameter with f-wave
symmetry is obtained. Possible relevance to real superconductors
is suggested.
\end{abstract}

\medskip


\begin{multicols}{2}
\narrowtext

\newpage
A fascination toward spin-triplet superconductivity 
has a long history, but 
recent experimental suggestions for triplet pairing
in a heavy fermion system UPt$_3$,\cite{Tou}  
organic conductors (TMTSF)$_2$X (X=ClO$_4$\cite{Lee},PF$_6$\cite{Lee2}),
and a ruthenate Sr$_2$RuO$_4$,\cite{Luke,Ishida} 
have renewed our interests in mechanisms of triplet 
superconductivity.  In particular, it is fairly intriguing
to investigate whether electron-electron repulsive interactions 
can lead to triplet superconductivity.
Ferromagnetic-spin-fluctuation mechanism 
has been proposed from the early days, but to our knowledge,  
realization of triplet 
superconductivity (at sizable temperatures) has yet to be established
theoretically for repulsive electron models 
with renormalization effects of the quasiparticles taken into account. 
The lifetime of the 
quasiparticles is important since this is a factor dominating $T_c$.

Recently, the present authors with Aoki have investigated    
the possibility of triplet pairing in the Hubbard model for a variety of 
lattice structures and band fillings using 
fluctuation exchange (FLEX) approximation.\cite{AKA}
A naive expectation is that triplet superconductivity may be realized 
when the band is away from half-filled and 
the density of states (DOS) at the Fermi level is large, since 
ferromagnetic fluctuations become strong in such a situation.
In ref.\cite{AKA}, however, it has turned out that 
the transition temperature $(T_c)$ of triplet superconductivity, if any,  
is too low to be detected as far as the cases surveyed there are concerned.
A typical case is a square lattice with
appreciable next nearest neighbor hoppings and dilute band fillings. 
First let us briefly review this situation as a reference for 
the results presented later.

\begin{figure}
\leavevmode\epsfysize=35mm \epsfbox{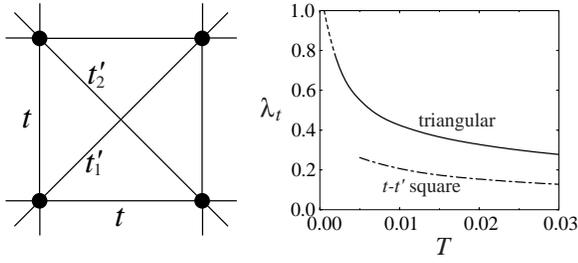}
\caption{The left panel shows the hopping integrals for square
$(t'_1=t'_2)$ or triangular $(t'_2=0)$ lattice. In the right panel,
$\lambda_t$ is plotted as a function of temperature 
for the Hubbard model on 
a square lattice with $t'_1=t'_2=0.4$, $U=6$, and $n=0.3$ 
(dash-dotted line),
or on an isotropic triangular lattice with $t'_1=1$, $U=8$, and $n=0.15$ 
(solid line).  In the latter case, a spline extrapolation to 
lower temperatures is also plotted (dashed line).
}
\label{fig1}
\end{figure}

We consider the Hubbard model, 
${\cal H}=\sum_{\langle i,j \rangle \sigma=\uparrow,\downarrow} 
t_{ij}c^{\dagger}_{i\sigma}c_{j\sigma}
+U\sum_i n_{i \uparrow}n_{i \downarrow},$
on a square lattice shown in Fig.\ref{fig1}, 
where $t(=1)$ is the nearest and $t'_1(=t'_2$ here) 
is the next nearest neighbor hopping.
In the FLEX calculation,\cite{Bickers} (i) Dyson's equation is 
solved to obtain the renormalized Green's function $G(k)$,
where $k\equiv({\bf k},i\epsilon_n)$ denotes the 2D wave-vectors and 
the Matsubara frequencies,
(ii) the effective electron-electron interaction $V^{(1)}(q)$ 
is calculated by collecting RPA-type bubbles and ladder diagrams consisting
of the renormalized Green's function, namely, 
by summing up powers of the irreducible susceptibility 
$\chi_{\rm irr}(q)\equiv -\frac{1}{N}\sum_k G(k+q)G(k)$ 
($N$:number of $k$-point meshes),
(iii) the self energy is obtained as 
$\Sigma(k)\equiv\frac{1}{N}\sum_{q} G(k-q)V^{(1)}(q)$, 
which is substituted into Dyson's equation in (i), 
and the self-consistent loops are repeated until convergence is attained.
Throughout the study, we take $64\times 64$ $k$-point meshes and 
the Matsubara frequencies $\epsilon_n$ from 
$-(2N_c-1)\pi T$ to $(2N_c-1)\pi T$ with $N_c$ up to 16384 in order to
ensure convergence at low temperatures. 

To obtain $T_c$, we solve the eigenvalue 
({\'E}liashberg) equation for the superconducting order parameter $\phi(k)$, 
\begin{eqnarray}
\lambda\phi(k)&=&-\frac{T}{N}
\sum_{k'}
\phi(k')|G(k')|^2 V^{(2)}(k-k'),
\label{eliash}
\end{eqnarray}
where the pairing interaction $V^{(2)}$, which mediates pair scattering from
$({\bf k,-k})$ to $({\bf k',-k'})[\equiv({\bf k+q,-k-q})]$, is given as 
\begin{eqnarray}
V_s^{(2)}(q)=
\frac{3}{2} \left[ \frac{U^2\chi_{\rm irr}(q)}{1-U\chi_{\rm irr}(q)} \right]
-\frac{1}{2} \left[ \frac{U^2\chi_{\rm irr}(q)}{1+U\chi_{\rm irr}(q)} \right]
\label{pair1}
\end{eqnarray}
for singlet pairing, and 
\begin{eqnarray}
V_t^{(2)}(q)=-\frac{1}{2} \left[ \frac{U^2\chi_{\rm irr}(q)}
{1-U\chi_{\rm irr}(q)} \right]
-\frac{1}{2} \left[ \frac{U^2\chi_{\rm irr}(q)}{1+U\chi_{\rm irr}(q)} \right] 
\label{pair2}
\end{eqnarray}
for triplet pairing.
In either case, the first (second) term represents the 
contribution from spin (charge) fluctuations. 
$T_c$ is the temperature at which the maximum eigenvalue 
$\lambda$ reaches unity.
We denote the eigenvalue and the order parameter 
for triplet (singlet) pairing as $\lambda_t$ 
($\lambda_s$) and $\phi_t$ $(\phi_s)$, respectively.

In ref.\cite{AKA}, $t'$, $U$, and the band filling $n$\cite{filling} 
were varied in search of triplet superconductivity,
but $\lambda_t$ remained below $\sim 0.2$ in the tractable 
temperature range as typically displayed in Fig.\ref{fig1} 
(dash-dotted line).
The main reason why triplet pairing instability is weak 
is because $|V_t^{(2)}|$ is only one third of $|V_s^{(2)}|$ 
when spin fluctuation is dominant as can be seen from eqs.(\ref{pair1}) and 
(\ref{pair2}).

\begin{figure}
\leavevmode\epsfysize=32mm \epsfbox{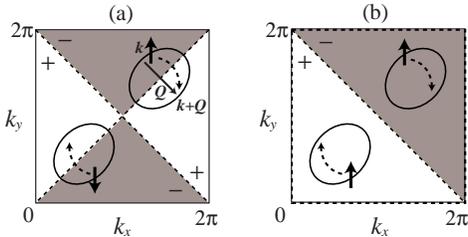}
\caption{Basic idea of the present mechanism is presented.
Intra-FS pair scattering (dashed curves) is mediated by $V^{(2)}({\bf Q})$
within each pieces of the disconnected FS (closed solid curves).
The dashed straight lines represent nodes of the order parameter 
for singlet (a) and triplet (b) pairing. 
}
\label{fig2}
\end{figure}

In this Letter, we propose that the above difficulty for 
spin-fluctuation-mediated triplet pairing in the Hubbard model 
can be overcome under certain conditions. Let us first present our idea.
We consider a situation (see Fig.\ref{fig2}) where (i) the 
Fermi surface (FS) is disconnected (preferably well separated) 
into two pieces which are located point symmetrically about ${\bf k=0}$, 
and (ii) the spin structure is pronounced 
around a wave vector {\bf Q} in such a way that two electrons 
with zero total momentum can be scattered {\it within each piece of the FS} 
(this process will be called intra-FS pair scattering hereafter)
by exchanging spin fluctuations having momentum ${\sim\bf Q}$.
Now, in order to have large $\lambda$, the quantity 
$-[\sum_{{\bf k,k'}\in FS} V^{(2)}({\bf k-k'})\phi({\bf k})\phi({\bf k'})]
/[\sum_{{\bf k}\in FS}\phi^2({\bf k})]$
(the numerator being $\sim -\sum_{{\bf k}\in FS} V^{(2)}({\bf Q})
\phi({\bf k})\phi({\bf k+Q})$) has to be positive and large. 
Then, pair scatterings from $({\bf k,-k})$ 
to $({\bf k+Q,-k-Q})$ for {\it singlet} pairing, 
mediated by {\it repulsive} $V_s^{(2)}({\bf Q})$, 
have to accompany a sign change in the order parameter
$\phi_s({\bf k})$ (Fig.\ref{fig2}(a)).
Hence the nodes of $\phi_s({\bf k})$ must intersect the FS.
For {\it triplet} pairing, by contrast, pairs can be scattered
within a region having the same sign in $\phi_t({\bf k})$
because $V_t^{(2)}({\bf Q})$ is {\it attractive}. 
In this case, since the gap nodes
(which exists due to triplet pairing symmetry
$\phi({\bf k})=-\phi(-{\bf k})$) do not intersect the FS
(Fig.\ref{fig2}(b)), the entire FS can be exploited for pairing, 
\begin{figure}
\leavevmode\epsfysize=100mm \epsfbox{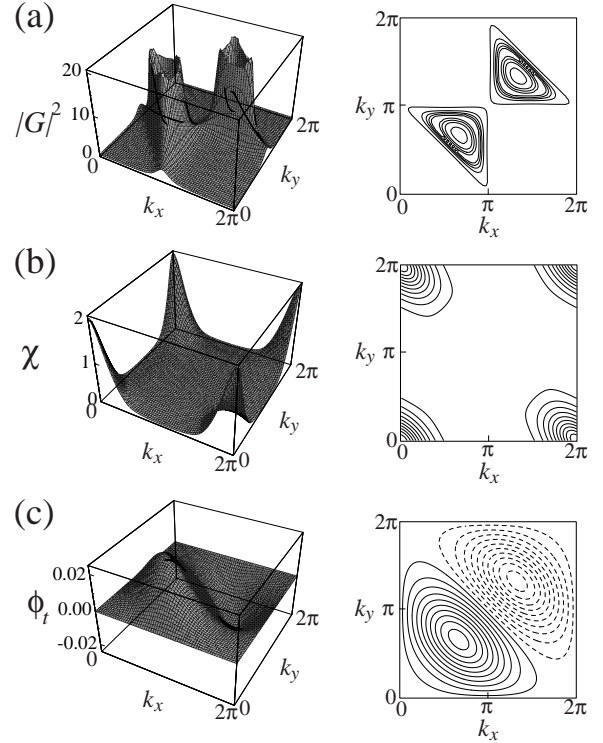}
\caption{$|G({\bf k},i\pi k_B T)|^2$ (a), $\chi({\bf k},0)$ (b), and 
$\phi_t({\bf k},i\pi k_B T)$ (c) are plotted for the Hubbard model on a 
triangular lattice with $t'_1=1$, $U=8$, $n=0.15$, and $T=0.01$. 
The right panels are contour-plots of the left panels.
}
\label{fig3}
\end{figure}
\noindent
so that triplet pairing may be enhanced. Quite recently,
a related proposal has been raised by Kohmoto and Sato 
for systems with both phonons and spin fluctuations present,\cite{KS} 
as discussed later.

The above conditions are not satisfied for the $t$-$t'$ square lattice 
because it has a connected FS.
As an example of a system in which the above conditions are indeed satisfied,
we consider the Hubbard model on an isotropic triangular lattice with
dilute band fillings. The band dispersion for $U=0$ is given by 
$\varepsilon_{\triangle}({\bf k})=2[\cos k_x +\cos k_y + \cos (k_x + k_y)]$
when we represent an isotropic triangular lattice by setting 
$t=t'_1=1$ \cite{t-sign} and $t'_2=0$ in Fig.\ref{fig1}.
Superconductivity on an isotropic 
triangular lattice has been studied by several authors,
\cite{Kino,Kondo,Vojta,KA} but their interest was mainly focused on $n\sim 1$.
In ref\cite{AKA}, possibility of triplet superconductivity was studied 
at quarter filling $(n=0.5)$, 
where ferromagnetic fluctuations become strong because 
the Fermi level lies right at the position where the DOS diverges.
However, $\lambda_t$ was again found to be small, which, in the present
context, is because the FS is connected.
If we set $n<0.5$, on the other hand, 
the FS is disconnected into two pieces, which are
centered respectively at $k=(2\pi/3,2\pi/3)$ and $k=(4\pi/3,4\pi/3)$.
Here we take $n=0.15$, where the two pieces of the FS are
well separated. In Fig.\ref{fig3},
we plot the FLEX result for 
$|G({\bf k},i\pi k_B T)|^2$ (a) and the spin susceptibility 
$\chi({\bf k},0)\equiv\chi_{\rm irr}({\bf k},0)/
[1-U\chi_{\rm irr}({\bf k},0)]$ (b)
against ${\bf k}$ for $U=8$ and $T=0.01$.
The FS as identified from the ridge in $|G({\bf k},i\pi k_B T)|^2$
is indeed disconnected into two pieces.
$\chi({\bf k},0)$ is sharply peaked at
${\bf k}= 0$ as seen in Fig.\ref{fig3}(b), indicating 
ferromagnetic fluctuations.\cite{chi-comment}
This is partially because the FS is small, but it is also 
because the Fermi level for $n=0.15$ is
still not too far away from the peak position of the DOS.
In this case, $\lambda_s$ is shown to be small, which is 
because $V_s^{(2)}({\bf Q})$ can only mediate pair scatterings 
in the vicinity of the nodes when ${\bf Q\sim 0}$.

If we turn to triplet pairing, 
the order parameter $\phi_t({\bf k},i\pi k_B T)$,  
plotted against ${\bf k}$ for $T=0.01$ in Fig.\ref{fig3}(c),
has f-wave (f$_{x^3-3xy^2}$-wave in the notation of the 
$C_{6}$ symmetry group) symmetry 
with three sets of nodal lines ($k_x\equiv 0 ({\rm mod} 2\pi), 
k_y\equiv 0,$ and $k_x+k_y\equiv 0$).
Comparing Figs.\ref{fig3}(a) and (c),
we can see that these nodes do not intersect the FS.
Accordingly, $\lambda_t$ (Fig.\ref{fig1},solid line) 
is strongly enhanced compared to the case for the $t$-$t'$ square lattice. 
A spline extrapolation to low temperatures suggests 
a possible low but finite $T_c$. 

\begin{figure}
\leavevmode\epsfysize=30mm \epsfbox{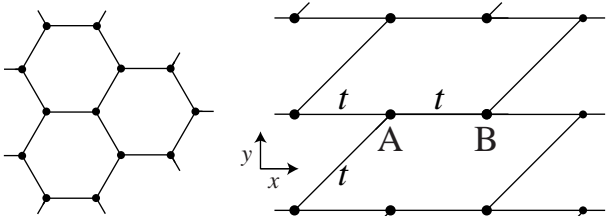}
\caption{For the honeycomb lattice shown in the left panel, we employ the 
topologically equivalent structure shown in the right.
}
\label{fig4}
\end{figure}

As another example, we next propose that the Hubbard model on a 
{\it honeycomb} lattice (Fig.\ref{fig4}) should also be interesting.
Since there are two sites (A and B) in a unit cell, 
this is a two-band system. The noninteracting band dispersion 
$\varepsilon_{\rm hc}({\bf k})=\pm\sqrt{\varepsilon_\triangle({\bf k})+3}$
has two pairs of vertex-sharing cones 
at $k=(2\pi/3,2\pi/3)$ and $k=(4\pi/3,4\pi/3)$, so 
again the FS becomes disconnected, this time for fillings close to $n=1$.

In the multiband version of FLEX,\cite{Koikegami,Kontani}  
the quantities $G$, $\chi$, $\Sigma$, and $\phi$ have $2\times 2$ matrix forms,
e.g., $G_{\alpha\beta}({\bf k}, i\omega_n)$, where $\alpha,\beta$ denote
A or B sites. The band representation of the Green's function and the 
order parameters is obtained by using the relation between the 
annihilation operators of upper $(u)$ and lower $(l)$ band electrons 
($c^u_{\bf k}$, $c^l_{\bf k}$) and those of A and B site electrons 
($c^{\rm A}_{\bf k}$, $c^{\rm B}_{\bf k}$). 
As for $\chi$, we diagonalize the 
$2\times 2$ matrix $\chi_{\alpha\beta}$ to obtain 
$\chi_{\pm}=(\chi_{\rm AA}+\chi_{\rm BB})/2\pm
\sqrt{[(\chi_{\rm AA}-\chi_{\rm BB})/2]^2+
|\chi_{\rm AB}|^2}$.

\begin{figure}
\leavevmode\epsfysize=140mm \epsfbox{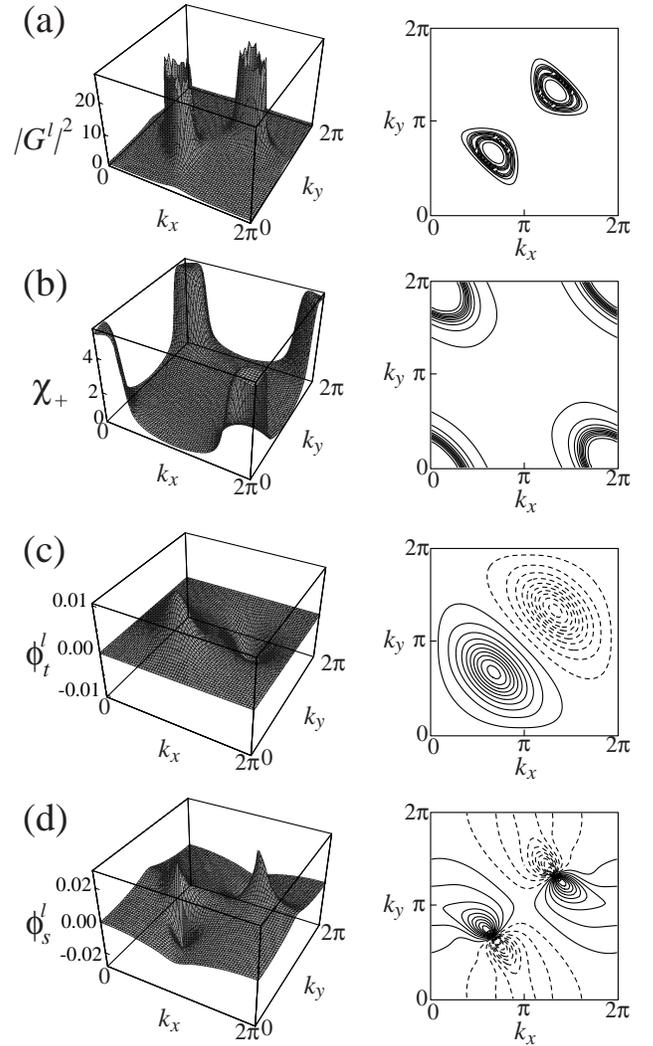}
\caption{$|G^l({\bf k},i\pi k_B T)|^2$ (a), $\chi_+({\bf k},0)$ (b), 
$\phi^l_t({\bf k},i\pi k_B T)$ (c), and $\phi^l_s({\bf k},i\pi k_B T)$ (d)
are plotted for the Hubbard model on a honeycomb lattice with
$U=8$, $n=0.95$, $\Delta_{\rm AB}=0$ and $T=0.01$.
}
\label{fig5}
\end{figure}

In Fig.\ref{fig5}, we plot $|G^l({\bf k},i\pi k_B T)|^2$ 
(a) and $\chi_{+}({\bf k},0)$ (b) for $n=0.95$,
$U=8$ and $T=0.01$. 
Since $\chi_{\rm AB}({\bf 0},0)$ is found to be negative, 
the peak around ${\bf k=0}$ in $\chi_+({\bf k},0)$ is an 
indication of antiferromagnetic fluctuations, as expected
for a nearly half-filled bipartite lattice system.
Note that $\chi_+({\bf k},0)$ has a broad structure
compared to the case for the triangular lattice (Fig.\ref{fig3}(b)). 

\begin{figure}
\leavevmode\epsfysize=40mm \epsfbox{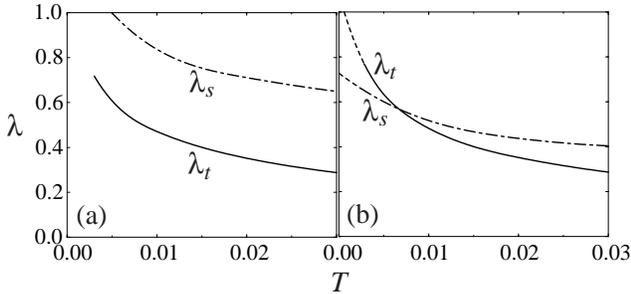}
\caption{$\lambda_t$ (solid line) and $\lambda_s$ (dash-dotted line) 
are plotted as functions of temperature 
for the Hubbard model on a honeycomb lattice with
$U=8$, $n=0.95$, and $\Delta_{\rm AB}=0$ (a) or $\Delta_{\rm AB}=5$ (b).
In (b), spline extrapolations to 
lower temperatures are indicated by dashed lines.
}
\label{fig6}
\end{figure}

If we turn to $\lambda_s$ and $\lambda_t$ as functions of $T$ in
Fig.\ref{fig6}(a), $\lambda_t$ is again large, 
but this time $\lambda_s$ is in fact larger.
We can trace this to the broad spin structure, 
for which spin fluctuations with relatively large momentum can be exchanged 
to mediate singlet pairing at wave vectors away from of the nodes.
Nevertheless, we can still observe that 
$\lambda_t$ is enhanced above $\lambda_s/3$
(recall that $|V_t^{(2)}|\sim |V_s^{(2)}|/3$), which should be 
due to the fact that the nodes in $\phi_s^l$ intersect the FS, 
while those in $\phi_t^l$ do not as seen 
by comparing Figs.\ref{fig5}(a) and (c)/(d).

We have found that $|\phi_{\rm AB}| >(<)\: |\phi_{\rm AA}|$ for singlet 
(triplet) pairing, meaning that singlet (triplet) pairing mainly takes 
place on different (same) sublattices. 
This is in fact consistent with the antiferromagnetic alignment of the spins.
Then, we can intuitively expect that triplet can dominate over singlet 
if we introduce a level offset, 
\noindent
$\Delta_{\rm AB}$, between A and B sites.
In Fig.\ref{fig6}(b), we plot $\lambda_s$ and $\lambda_t$ 
for the same parameter values as in Fig.\ref{fig6}(a), 
except for a finite  $\Delta_{\rm AB}=5$. Triplet 
pairing indeed dominates over singlet pairing at low temperatures, 
and here again a possible finite $T_c$ for triplet superconductivity 
is suggested. The intuitive picture can be paraphrased 
in the momentum space that the peak structure of  $\chi$ 
(not shown) becomes sharper when $\Delta_{\rm AB}\neq 0$. 

Finally, let us make some remarks concerning possible relevance to 
real superconductors. As for UPt$_3$, if we examine the FS 
calculated by FLAPW method,\cite{Kimura} 
we notice that there are two disconnected pockets 
(band 37 in ref.\cite{Kimura}),\cite{BZ} which, 
in our view, is favorable for triplet pairing. 
It would be an interesting future problem to investigate in detail 
the applicability of the present mechanism.

Disconnected FS can arise similarly 
in graphite intercalation compounds (GIC),
except that the FS is cylindrical (quasi 2D).\cite{Inoshita} 
This is because graphite is a system where 
honeycomb sheets of carbon atoms are stacked.
Although spin fluctuations in GIC may not be strong enough to induce 
superconductivity purely electronically,
the disconnectivity of the FS itself should be favorable 
for triplet pairing, so a cooperation between 
certain phonon modes and (weak) spin 
fluctuations might lead to triplet superconductivity (even in the 
absence of $\Delta_{\rm AB}$ considered above).
Namely, if attractive intra-FS pair scatterings mediated by phonons are 
present, antiferromagnetic spin fluctuations as considered here 
would work constructively with phonons to enhance intra-FS 
pair scatterings for triplet pairing, while the converse is true for 
singlet pairing. Experimentally, 
although triplet pairing has not
been claimed to our knowledge, 
a large value of $H_{c2}$ (extrapolated to $T=0$) observed 
in C$_4$KTl$_{1.5}$\cite{Iye} is in fact reminiscent of a large 
H$_{c2}$ in (TMTSF)$_2$X.\cite{Lee,Lee2}

As for (TMTSF)$_2$X and  Sr$_2$RuO$_4$, 
Kohmoto and Sato have recently proposed that 
disconnectivity (quasi-one dimensionality) of the FS, 
along with the presence of spin fluctuations originating from 
the nesting of the FS, plays an essential role in stabilizing 
{\it phonon-mediated} triplet p-wave pairing.\cite{KS} 
Our study is related to this proposal in that disconnectivity is important,
but in these systems, as seen in Kohmoto and Sato's argument,
the  dominant spin fluctuations have wave-vectors that {\it bridge} 
the two pieces of the FS, so that they mediate {\it inter}-FS 
pair scatterings\cite{KS-comment} rather than {\it intra}-FS ones. 
Thus, our {\it purely} spin-fluctuation-mediated pairing mechanism 
does not directly apply to these materials, although we do believe 
that the disconnectivity of the FS may be playing a certain role
in realizing triplet superconductivity.

We wish to thank Prof. H. Aoki for illuminating discussions.
We also thank Prof. M. Kohmoto and Dr. M. Sato for fruitful discussions and 
sending us ref.\onlinecite{KS} prior to publication.
K.K. thanks Dr. H. Kontani for discussions on multiband FLEX.
Numerical calculations were performed at the Computer Center,
University of Tokyo, and at the Supercomputer Center,
ISSP, University of Tokyo.
K.K. acknowledges support by the Grant-in-Aid for Scientific
Research from the Ministry of Education of Japan, while 
R.A. acknowledges support by the JSPS Research Fellowships for
Young Scientists.



\end{multicols}

\begin{references}
\bibitem{filling} Throughout the paper, we define the band filling as 
$n=$[number of electrons]/[number of sites].
\bibitem{t-sign} Since a triangular lattice is non-bipartite, the sign
of $t$ is important. Note that we are taking $t>0$ and $n<1$.
\bibitem{chi-comment} The maximum value of $\chi$ stays around 
a moderate value $(\sim 2)$ down to the lowest temperature considered 
($T=0.002$). The situation is similar for the honeycomb lattice.
This is in contrast with the case for the $t$-$t'$ square lattice, 
where Max[$\chi$] exceeds 10 at $T=0.01$, and 
continues to grow for lower temperatures.
\bibitem{BZ} Usually the Brillouin zone (BZ) of UPt$_3$ 
is taken to be a hexagonal prism,
where six pieces of FS are located at the edges of the BZ. 
The two pockets are more noticeable if we  
take a rhombic prism as the BZ. The same applies to GIC.
\bibitem{KS-comment} 
In inter-FS pair scatterings, electrons are scattered from one piece of 
the FS to the other. Such scatterings mediated by spin fluctuations 
suppress p-wave pairing,
but they also suppress s-wave pairing even more strongly, 
which is the reason why p-wave dominates over s-wave 
in the presence of phonons in Kohmoto-Sato's mechanism.
\end{references}
\end{document}